\documentclass[12pt, pdftex]{article}
\usepackage[]{graphicx}
\usepackage[]{color}
\makeatletter
\def\maxwidth{ %
  \ifdim\Gin@nat@width>\linewidth
    \linewidth
  \else
    \Gin@nat@width
  \fi
}
\makeatother

\definecolor{fgcolor}{rgb}{0.345, 0.345, 0.345}

\usepackage{framed}
\makeatletter
 {\par\unskip\endMakeFramed%
 \at@end@of@kframe}
\makeatother

\definecolor{shadecolor}{rgb}{.97, .97, .97}
\definecolor{messagecolor}{rgb}{0, 0, 0}
\definecolor{warningcolor}{rgb}{1, 0, 1}
\definecolor{errorcolor}{rgb}{1, 0, 0}

\usepackage{alltt}

\usepackage[a4paper]{geometry}

\usepackage[round]{natbib}
\usepackage{url}
\usepackage[colorlinks,citecolor=blue,urlcolor=blue,linkcolor=blue]{hyperref}
\usepackage[T1]{fontenc}              
\usepackage[utf8]{inputenc}         
\usepackage[english]{babel}
\usepackage{doi}

\usepackage{amsthm,amsmath,amsfonts,amssymb} 
\usepackage{amsthm}                   
\usepackage{graphicx}                 
\usepackage{subfig}                   
\usepackage{lscape}
\usepackage{booktabs}                         
\usepackage{listings}                 
\usepackage{tikz}
\usepackage{hyperref}
\usepackage{algorithm2e}
\usepackage{mathtools}
\usepackage{enumerate}

\usepackage{relsize}
\usepackage{authblk}
\usepackage{epstopdf}
          
\usepackage{array}
\usepackage{amssymb}

\newcommand{\blanco}[1]{  } 

\DeclareRobustCommand\xdot{\futurelet\token\Xdot}
\def\Xdot{\ifx\token\bgroup.\else\ifx\token\egroup.\else
  \ifx\token\/.\else\ifx\token\ .\else\ifx\token!.\else
  \ifx\token,.\else\ifx\token:.\else\ifx\token;.\else
  \ifx\token?.\else\ifx\token/.\else\ifx\token'.\else
  \ifx\token).\else\ifx\token-.\else\ifx\token+.\else
  \ifx\token~.\else
  \ifx\token.\else.\ \fi\fi\fi\fi\fi\fi\fi\fi\fi\fi\fi\fi\fi\fi\fi\fi}

\newcolumntype{L}[1]{>{$}p{#1}<{$}} 
\newcolumntype{M}{>{$}l<{$}}    
\newcolumntype{N}{>{$}c<{$}}    

{\section*{Acknowledgments}}%
{}

\DeclareMathOperator{\Nor}{N} 
  
\DeclareMathOperator{\Unif}{U} 

\newcommand{\ml}[2][1]{
\ifthenelse{#1 = 1}%
 {\hat{#2}_{\scriptscriptstyle{\mathrm{ML}}}}%
 {\hat{#2}^{#1}_{\scriptscriptstyle{\mathrm{ML}}}}
}

\newcommand{\given}{\,\vert\,} 

\newcommand{\partialv}[3][]{%
  \ifthenelse{\isempty{#1}}
  {\frac{\partial\,#2}{\partial\,#3}}
  {\frac{\partial^{#1} #2}{\partial\,#3^{#1}}} 
} 

\newcommand{\partials}[3][]{%
  \ifthenelse{\isempty{#1}}
  {\frac{d\,#2}{d\,#3}}
  {\frac{d^{#1} #2}{d\,#3^{#1}}}
} 

\newcommand{\dseps}[2][]{%
  \ifthenelse{\isempty{#1}}
  {\frac{d}{d#2}}
  {\frac{d^{#1}}{d#2^{#1}}}
}

\newcommand{\dsepv}[2][]{%
  \ifthenelse{\isempty{#1}}
  {\frac{\partial\,}{\partial\,#2}}
  {\frac{\partial^{#1}}{\partial\,#2^{#1}}}
}

\newcommand{\mailto}[2][]{%
  \ifthenelse{\isempty{#1}}%
  {\href{mailto:#2}{\nolinkurl{#2}}}%
  {\href{mailto:#2}{\nolinkurl{#1}}}%
}

\IfFileExists{upquote.sty}{\usepackage{upquote}}{}
\begin{document}

\title{Fast and accurate Bayesian model criticism and conflict diagnostics using R-INLA}
\author[1]{Egil Ferkingstad\thanks{\url{egil.ferkingstad@gmail.com}}}
\author[2]{Leonhard Held}
\author[3]{H{\aa}vard Rue}
\affil[1]{Science Institute, University of Iceland, Reykjavik, Iceland}
\affil[2]{Epidemiology, Biostatistics and Prevention Institute, University of Zurich, Switzerland}
\affil[3]{CEMSE Division, King Abdullah University of Science and Technology (KAUST),
  Saudi Arabia}
\date{\today}
\maketitle

\begin{abstract}
Bayesian hierarchical models are increasingly popular for realistic
modelling and analysis of complex data. This trend is accompanied by
the need for flexible, general and computationally efficient methods
for model criticism and conflict detection. Usually, a Bayesian
hierarchical model incorporates a grouping of the individual
data points, as, for example, with individuals in repeated
measurement data. In such cases, the following question
arises: Are any of the groups ``outliers,'' or in conflict
with the remaining groups? Existing general approaches
aiming to answer such questions tend to be extremely
computationally demanding when model fitting is based
on Markov chain Monte Carlo. We show how group-level
model criticism and conflict detection can be carried out
quickly and accurately through integrated nested
Laplace approximations (INLA). The new method is implemented
as a part of the open-source R-INLA package for Bayesian computing
(\url{http://r-inla.org}).\\
\smallskip
\noindent \textbf{Keywords:} Bayesian computing; Bayesian modelling;
INLA; model criticism; model assessment; latent Gaussian models
\end{abstract}
\section{Introduction}
\label{sec:intro}
The Bayesian approach gives great flexibility for realistic modelling of complex data. However, any assumed model
may be unreasonable in light of the observed data, and different parts of the data may be in conflict with each other.
Therefore, there is an increasing need for flexible, general and computationally efficient methods for model criticism
and conflict detection. Usually, a Bayesian hierarchical model incorporates a grouping of the individual data points. For
example, in clinical trials, repeated measurements are grouped by patient; in disease mapping, cancer counts may be
grouped by year or by geographical area. In such cases, the following question arises: Are any of the groups ``outliers,''
or in conflict with the remaining groups? Existing general approaches aiming to answer such questions tend to be
extremely computationally demanding when model fitting is based on Markov chain Monte Carlo (MCMC). We show
how group-level model criticism and conflict detection can be carried out quickly and accurately through integrated
nested Laplace approximations (INLA). The new method is implemented as a part of the open-source R-INLA package
for Bayesian computing
(\url{http://r-inla.org}); see Section \ref{sec:impl-r-inla} for
details about the implementation.

The method is based on so-called \emph{node-splitting}~\citep{MR2312289,presanis2013}, which has previously been used
by one of the authors~\citep{sauter-held-2015} in 
\emph{network meta-analysis}~\citep{lumley2002,dias.etal2010} to
investigate whether direct evidence on treatment comparisons differs
from the evidence obtained from indirect comparisons only. The assessment of possible network
inconsistency by node-splitting implies that a network meta-analysis
model is fitted repeatedly for every directly observed comparison of
two treatments where also indirect evidence is available. For network
meta-analysis, the INLA
approach has been recently shown to provide a fast and reliable
alternative to node-splitting based on MCMC
\citep{sauter-held-2015}. This has inspired the work described here
to develop general INLA routines for model criticism in Bayesian
hierarchical models.

This paper is organized as follows. In Section \ref{sec:method}, we
describe the methodology based on the latent Gaussian modelling framework
outlined in Section \ref{sec:inla}. Specifically, Section \ref{sec:model-criticism} describes 
general approaches to model criticism in Bayesian hierarchical models, while 
Section \ref{sec:group-split-inla} outlines our proposed method for routine analysis
within INLA. Section \ref{sec:impl-r-inla} discusses the
implementation in the R-INLA package. We provide two applications in Section
\ref{sec:applications} and close with some discussion in Section \ref{sec:disc}.

\section{Methodology}
\label{sec:method}

\subsection{Latent Gaussian models and INLA}
\label{sec:inla}

The class of \emph{latent Gaussian models} (LGMs) includes many Bayesian hierarchical models of interest. This model class is intended to provide a good trade-off between flexibility and computational efficiency: for models within the LGM class, we can bypass MCMC completely and obtain accurate deterministic approximations using INLA. The LGM/INLA framework was introduced in~\citet{rue2009}; for an up-to-date review covering recent developments, see~\citet{rue2017}.

We shall only give a very brief review of LGM/INLA here; for more details, see the aforementioned references, as
well as~\citet{rue2005} for more background on Gaussian Markov random fields, which form the computational
backbone of the method. The basic idea of the LGM is to split into a hierarchy with three levels:
\begin{enumerate}
\item The \emph{(hyper)prior level}, with the parameter vector
  $\boldsymbol{\theta}$. It is assumed that the dimension of
  $\boldsymbol{\theta}$ is not too large (less than 20; usually 2 to
  5). Arbitrary priors (with restrictions only on smoothness /
  regularity conditions) can be specified independently for each
  element of  $\boldsymbol{\theta}$ (and there is ongoing work on
  allowing for more general joint priors; see~\citet{simpson2017} for
  ideas in this direction). Also, it is possible to use the INLA-estimated posterior distribution of $\boldsymbol{\theta}$ as the prior for a new run of INLA; we shall use this feature here. 
\item The \emph{latent level}, where the \emph{latent field} $\boldsymbol{x}$ is assumed to have a multivariate Gaussian distribution conditionally on the hyperparameters $\boldsymbol{\theta}$. The dimension of $\boldsymbol{x}$ may be very large, but conditional independence properties of $\boldsymbol{x}|\boldsymbol{\theta}$ typically implies that the precision matrix (inverse covariance matrix) in the multivariate Gaussian is sparse (i.e.~$\boldsymbol{x}$ is a Gaussian Markov random field), allowing for a high level of computational efficiency using methods from sparse linear algebra.
\item The \emph{data level}, where observations $\boldsymbol{y}$ are conditionally independent given the latent field $\boldsymbol{x}$. In principle, the INLA methodology allows for arbitrary data distributions (likelihood functions); a long list of likelihood functions (including normal, Poisson, binomial and negative binomial) are currently implemented in the R-INLA package.
\end{enumerate}
It is assumed that the joint posterior of $(\boldsymbol{x}, \boldsymbol{\theta})$ can be factorized as
\begin{equation}
\pi(\boldsymbol{x}, \boldsymbol{\theta}|\boldsymbol{y}) \propto \pi(\boldsymbol{\theta}) \pi(\boldsymbol{x}|\boldsymbol{\theta})
\prod_{i \in \mathcal{I}} \pi(y_i | x_i, \boldsymbol{\theta})
\label{eq:lgm}
\end{equation}
where $\mathcal{I}$ is a subset of $\{1,2,\ldots,n\}$ and where $n$ is the dimension of the latent field $\boldsymbol{x}$. 

In R-INLA, the model is defined in a format generalizing the well-known generalized linear/additive (mixed) models: for each $i$, the \emph{linear predictor} $\eta_i$ is modelled using additive building blocks, for example as (Equation (2) of~\citet{rue2017}):
\begin{equation}
  \eta_i = \mu + \sum_j \beta_j z_{ij} + \sum_k f_{k, j_k(i)}
  \label{eq:linpred}
\end{equation}
where $\mu$ is an overall intercept, the $\beta_j$ are ``fixed effects'' of covariates $z_{ij}$, and the $f_k$ terms represent Gaussian process model components, which can be used for random effects models, spatial models, time-series models, measurement error models, spline models and so on. The wide variety of possible $f_k$ terms makes this set-up very general, and it covers most of the Bayesian hierarchical models used in practice today.
As in the usual generalized linear model framework~\citep{mccullagh1989}, the linear predictor $\eta_i$ is related to the mean of data point $y_i$ (or, more generally, the location parameter of the likelihood) through a known link function $g$. For example, in the Poisson likelihood with expected value $\lambda_i$, the log link function is used, and the linear predictor is $\eta_i = \log(\lambda_i)$. Assuming {\it a priori} independent model components and a multivariate Gaussian prior on $(\mu, \boldsymbol{\beta})$, we can define the latent field as
$\boldsymbol{x} = (\boldsymbol{\eta}, \mu, \boldsymbol{\beta}, \boldsymbol{f}_1,  \boldsymbol{f}_2, \ldots)$, with a small number of hyperparameters $\boldsymbol{\theta}$ arising from the likelihood and model components. It follows that  $\boldsymbol{x}$ has a multivariate Gaussian distribution conditionally on $\boldsymbol{\theta}$, so we are in the LGM class described earlier. For advice on how to set the priors on $\boldsymbol{\theta}$, see~\citet{simpson2017}.

Why restrict to the LGM class? The reason is that quick, deterministic
estimation of posterior distributions is possible for LGMs, using INLA. Here, we only give a brief sketch of how this works; for further details, see~\citet{rue2009} and \citet{rue2017}. Assume that the objects of interest are the marginal posterior distributions $\pi(x_i|\boldsymbol{y})$ and $\pi(\theta_j|\boldsymbol{y})$ of the latent field and hyperparameters, respectively. The full joint distribution is 
\[
\pi(\boldsymbol{x}, \boldsymbol{\theta}, \boldsymbol{y}) = \pi(\boldsymbol{\theta}) \pi(\boldsymbol{x}|\boldsymbol{\theta})
\prod_{i \in \mathcal{I}} \pi(y_i | x_i, \boldsymbol{\theta})
\]
(cf.~Equation~\eqref{eq:lgm} earlier). We begin by noting that a Gaussian approximation $\tilde \pi_G(\boldsymbol{x} | \boldsymbol{\theta}, \boldsymbol{y})$ can be calculated relatively easily, by matching the mode and curvature of the mode of $\pi(\boldsymbol{x} | \boldsymbol{\theta}, \boldsymbol{y})$.
Then, the
Laplace approximation~\citep{tierney1986} of the posterior $\pi(\boldsymbol{\theta}|\boldsymbol{y})$ is given by
\[
\tilde \pi(\boldsymbol{\theta}|\boldsymbol{y}) \propto \left.\frac{\pi(\boldsymbol{x}, \boldsymbol{\theta}, \boldsymbol{y})}{\tilde \pi_G(\boldsymbol{x} | \boldsymbol{\theta}, \boldsymbol{y})}\right\vert_{ \boldsymbol{x}= \boldsymbol{\mu} (\boldsymbol{\theta)}},
\]
where $\boldsymbol{\mu} (\boldsymbol{\theta})$ is the mean of the Gaussian approximation. Because the dimension of $\boldsymbol{\theta}$ is not too large, the desired marginal posteriors can then be calculated using numerical integration:
\begin{eqnarray*}
   \tilde \pi(\theta_j|\boldsymbol{y}) &=& \int \tilde \pi(\boldsymbol{\theta}| \boldsymbol{y}) \mathrm{d}\boldsymbol{\theta}_{-j},\nonumber\\
   \tilde \pi(x_i|\boldsymbol{y}) &=& \int \tilde \pi(x_i|\boldsymbol{\theta},\boldsymbol{y}) \tilde \pi(\boldsymbol{\theta}| \boldsymbol{y}) \mathrm{d}\boldsymbol{\theta},
\end{eqnarray*}
where the
approximation $\pi(x_i|\boldsymbol{\theta},\boldsymbol{y})$ is
calculated using a skew normal density~\citep{azzalini1999}.
This approach is
accurate enough for most purposes, but see~\citet{ferkingstad2015} for
an improved approximation useful in particularly difficult cases. 
 
The methodology is implemented in the open-source R-INLA R package;
see \url{http://r-inla.org} for documentation, case studies
etc. There are also a number of features allowing to move beyond the
framework above (including  the ``linear combination'' feature
we will use later), see \citet{martins2013}.

\subsection{Model criticism and node-splitting}
\label{sec:model-criticism}

Using INLA (or other Bayesian computational approaches), we can fit a wide range
of possible models. With this flexibility comes the need to
check that the model is adequate in light of the observed data. In the
present paper, we will not discuss model \emph{comparison}, except for
pointing out that R-INLA already has implemented standard model
comparison tools such as the deviance information criterion~\citep{spiegelhalter2002} and
then Watanabe-Akaike information criterion~\citep{watanabe2010, vehtari2016}. We shall only discuss
model~\emph{criticism}, that is, investigating model adequacy without
reference to any alternative model.

Popular approaches to Bayesian model criticism include posterior
predictive $p$-values~\citep{gelman1996} as well as cross-validatory
predictive checks such as the conditional predictive ordinate
(CPO)~\citep{pettit1990} and the cross-validated probability integral
transform (PIT)~\citep{dawid1984}. All of these measures are easily
available from R-INLA (and can be well approximated without actually
running a full cross-validation); see~\citet{held2010} for a review and practical
guide. As noted by~\citet{held2010} (see also~\citet{bayarri2007}),
posterior predictive $p$-values have the drawback of not being properly
calibrated: they are not uniformly distributed, even if the data come
from the assumed model, and are therefore difficult to interpret. The
problem comes from ``using the data twice,'' as the same data are used
both to fit and criticize the model. 
Because cross-validation avoids this pitfall, we recommend the use of cross-validatory
measures such as CPO or PIT.

For illustration consider the following simple example. Suppose $y_1,
\ldots, y_n$ are independent realisations from a $\Nor(\mu, \sigma^2)$
distribution with unknown $\mu$ and known $\sigma^2$. For simplicity
suppose a flat prior is used for $\mu$.  Then:
\begin{eqnarray}
\mu \given y_{-i} & \sim & \Nor(\bar{y}_{-i}, \sigma^2/(n-1)) \label{eq:eq1} \\
\mu \given y_i & \sim & \Nor(y_i, \sigma^2) \label{eq:eq2} 
\end{eqnarray}
so
\[
y_i \given y_{-i} = \int [y_i \given \mu] [\mu \given y_{-i}] d\mu \sim \Nor(\bar{y}_{-i}, \underbrace{\sigma^2 \cdot n/(n-1)}_{ =: \tilde{\sigma}^2}) 
\]
and we obtain the PIT value
\begin{equation}\label{eq:PIT}
\mbox{PIT}_i = \Pr\{ Y_i \leq y_i \given y_{-i}\} = \Phi((y_i - \bar{y}_{-i})/\tilde{\sigma}).
\end{equation}
If the model is true, then $(Y_i - \bar{Y}_{-i})/\tilde{\sigma} \sim 
\Nor(0,1)$ and therefore $\mbox{PIT}_i \sim \Unif(0,1)$.

Alternatively we could also consider the difference of \eqref{eq:eq1} and \eqref{eq:eq2}
\[
\delta = \mu \given y_i - \mu \given y_{-i} \sim \Nor(\underbrace{\bar{y}_{-i} - {y}_{i}}_{\mu_{\delta}}, \underbrace{\sigma^2 + \sigma^2/(n-1)}_{ = \tilde{\sigma}^2})
\]
 and compute with $\tilde{\delta} = (\delta - \mu_{\delta})/\tilde{\sigma} \sim \Nor(0,1)$ the tail probability 
\begin{equation}\label{eq:latent}
\Pr\{ \delta \leq 0  \} = \Pr\{ \tilde{\delta} \leq  - \mu_{\delta}/\tilde{\sigma} \} = \Phi((y_i - \bar{y}_{-i})/\tilde{\sigma}),
\end{equation}
which is the same as \eqref{eq:PIT}. This shows that in this simple
model, the PIT approach based on observables ($y_i$) is equivalent to
the latent approach based on the distribution of $\delta$.

Both \eqref{eq:PIT} and \eqref{eq:latent} are one-sided tail
probabilities. A two-sided version of \eqref{eq:latent} can easily be
obtained by considering $\tilde{\delta}^2  \sim \chi^2(1)$ so the two-sided
tail probability is
\begin{eqnarray*}
p & = & 2 \min \{\Pr\{ \tilde{\delta} \leq  - \mu_{\delta}/\tilde{\sigma} \}, 
1 - \Pr\{ \tilde{\delta} \leq  - \mu_{\delta}/\tilde{\sigma} \}  \} \\
& = & 1 - \Pr\{{\tilde{\delta}^2} \leq \mu_{\delta}^2/\tilde{\sigma}^2  \} \\
& = & \Pr\{{\tilde{\delta}^2} \geq \mu_{\delta}^2/\tilde{\sigma}^2  \}.
\end{eqnarray*}

A limitation of all the measures discussed above is that they only
operate on the data level and on a point-by-point basis: each data
point $y_i$ is compared with the model fit either from the remaining
data points $\boldsymbol{y}_{-i}$ (for the cross-validatory measures)
or from the full data $\boldsymbol{y}$ (for posterior predictive
measures). However, often there is an interest in
queries of model adequacy where we look more generally into whether
any sources of evidence conflict with the
assumed model. As a simple example, consider a medical study with
repeated measures: Each patient $j=1,\ldots,J$ has some measurement
taken at time
points $t=1,\ldots,T$, giving rise to data points $z_{jt}$
(corresponding to the ``$y_i$'' in the general notation). For this
example, it seems natural to consider each patient as a ``source of
evidence,'' and we may wish to consider model adequacy \emph{for each
  patient}, for example, asking whether any particular patient is an
``outlier'' according to the model. However, the methods above do not
provide any easy way of answering such questions, as they are limited
to looking directly at the data points $z_{jt}$.
To answer questions such as ``is patient NN an outlier,'' we need to  approach the model criticism task somewhat
differently: rather than looking directly at data points $y_i$, we
need to lift the model criticism up to the latent level of the
model. 

In our view, the most promising approach aiming to achieve this goal
is the cross-validatory method known as \emph{node-splitting},
proposed by~\citet{MR2312289} and extended by~\citet{presanis2013}. 
The word ``node'' refers to a vertex
(node) in a directed acyclic graph (DAG) representing the model, but for
our purposes here, it is not necessary to consider the DAG for the
model (``parameter-splitting'' would perhaps be a better name for
the method). To explain node-splitting, we begin by considering some
grouping of the data. For example, in the repeated measures case described
above, the data $\boldsymbol{y}$ can be grouped by patient
$j=1,\ldots,J$, such that $\boldsymbol{y}_j = (y_{j1}, \ldots,
y_{jT})'$. In general, we let $\boldsymbol{y}_j$ denote the data for
group $j$, while $\boldsymbol{y}_{-j}$ are the data for all remaining
groups $1,\ldots,j-1,j+1,\ldots,J$. Further, consider 
a (possibly multivariate) group-level parameter
$\boldsymbol{\gamma}_j$ in the model. (For the moment,
$\boldsymbol{\gamma}_j$ may be any parameter of the model, but we shall
make a particular choice of $\boldsymbol{\gamma}_j$ in the next
section.) The basic idea behind
node-splitting is to consider both the ``within-group'' posterior 
$\pi(\boldsymbol{\gamma}_j | \boldsymbol{y}_j)$ and the
``between-group'' posterior $\pi(\boldsymbol{\gamma}_j |
\boldsymbol{y}_{-j})$ for each group $j$, in effect splitting the evidence informing the ``node'' $\boldsymbol{\gamma}_j$ 
between 
\begin{enumerate}
\item  the evidence provided by group $j$, and
\item the evidence provided by the other groups
\end{enumerate}
Now, the intuition is that
if group $j$ is consistent with the other groups, then
$\pi(\boldsymbol{\gamma}_j | \boldsymbol{y}_j)$ and $\pi(\boldsymbol{\gamma}_j |
\boldsymbol{y}_{-j})$ should not differ too much. A test for consistency between the groups can be constructed by considering the difference
\begin{equation}
\boldsymbol{\delta}_j = \boldsymbol{\gamma}_j^{(-j)} -
\boldsymbol{\gamma}_j^{(j)},
\label{eq:delta1}
\end{equation}
where $\boldsymbol{\gamma}_j^{(-j)}
\sim \pi(\boldsymbol{\gamma}_j | \boldsymbol{y}_{-j})$ and $\boldsymbol{\gamma}_j^{(j)}
\sim \pi(\boldsymbol{\gamma}_j | \boldsymbol{y}_j)$, and testing $H_0$:
$\boldsymbol{\delta}_j = \boldsymbol{0}$ for each group $j$. We shall explain a specific
way to
construct such a test in Section~\ref{sec:group-split-inla}, but
see~\citet{presanis2013} for further details and alternative
approaches to performing the test for conflict detection. 

\subsection{General group-wise node-splitting using R-INLA}
\label{sec:group-split-inla}

In principle, the node-splitting procedure described in
Section~\ref{sec:model-criticism} can be implemented using sampling-based (i.e.~MCMC-based) methods; see, for example,~\citet{MR2312289} and
\citet{presanis2013}. However, there are a number of problems with
existing approaches, and these concerns are both of a theoretical and
practical nature:
\begin{enumerate}
\item In our view, the most important problem is computational:
  running the group-wise node-splitting procedure using MCMC becomes very computationally
  expensive for all but the smallest models and data sets.
\item Related to the computational problem above is the
  implementational issue: even though there exist well-developed
  computational engines for MCMC (such as OpenBUGS \citep{lunn2009},
  JAGS \citep{plummer2016}
  and STAN \citep{carpenter2017}), there
  is not to our knowledge any available software that can add the
  node-split group-wise model criticism functionality to an existing
  model. Thus, for a given model, actually implementing the node-split
  requires a non-trivial coding task, even if the model itself is
  already implemented and running in OpenBUGS, JAGS or STAN.
\item There is also an unsolved problem concerning what should be done with shared
  parameters. For example, if the node-split parameter corresponds to 
  independent random effects $\gamma_j \sim N(0, \sigma^2)$, the
  variance $\sigma^2$ is then shared between the different
  $\gamma_j$. The existence of the shared parameter $\sigma^2$ implies
  that $\boldsymbol{\gamma}_j^{(-j)}$ and
  $\boldsymbol{\gamma}_j^{(j)}$ from Equation~\eqref{eq:delta1} are not
  independent, and there are also concerns about estimating $\sigma^2$
  based on group $j$ alone, since the number of data points within
  each point may be small. \citet{MR2312289} and \citet{presanis2013}
  suggest to treat shared parameters as ``nuisance parameters'', by
  placing a ``cut'' in DAG, preventing information flow from group $j$
  to the shared parameter. However, it turns out that implementing the
  ``cut'' concept within MCMC algorithms is a difficult matter. 
  Discussion of practical and conceptual issues connected to
  the ``cut'' concept has mainly been taking place on
  blogs and mailing lists rather than in traditional refereed
  journals, but a thorough review is nevertheless provided
  by~\citet{plummer2015}, who is the author of the JAGS
  software. \citet{plummer2015} shows that the ``cut'' functionality
  (recommended by~\citet{presanis2013})
  implemented in the OpenBUGS software ``does not converge to a
  well-defined limiting distribution,'' i.e.~it does not work. We are
  not aware of any correct general MCMC implementation. 
\end{enumerate}

 Our approach using INLA solves all of the issues above. The first
 issue (computation intractability) is dealt with simply because INLA
 is  orders of magnitude faster than MCMC, making cross-validation-based
 approaches manageable. The second issue is also dealt with: We provide
 a general implementation that only needs the model specification and
 an index specifying the grouping, so implementation of the extra
 step of running the node-split model checking (given that the user already has a working
 INLA model) is a trivial task. We also provide an elegant solution of
 the third issue of ``shared parameters''/''cuts,'' implementing the
 ``cut'' 
 using an approach equivalent to
 what~\citet{plummer2015} calls ``multiple imputation.''

The key to solving the second issue, that is, making a general,
user-friendly implementation, is to make one specific choice of
parameter for the node-split: the linear predictor
$\boldsymbol{\eta}$. Intuitively, this is what the model ``predicts''
on the latent level, so it seems to make sense to use
it for a measure of predictive accuracy. The main advantage of
$\boldsymbol{\eta}$ is that it will always exist for any model in
INLA, making it possible to provide a general implementation. Thus,
the node split is implemented by providing a grouping $j=1,\ldots,J$.
Formally, the grouping is a partition of the index set of the
observed data $\boldsymbol{y}$. For each group $j$, we collect the
corresponding elements in $\boldsymbol{\eta}$ (i.e.~corresponding to data
$\boldsymbol{y}_j$ contained in group $j$) into the vector
$\boldsymbol{\eta}_j$. The node-splitting procedure sketched at the
end of Section~\ref{sec:model-criticism} then follows through, simply
by replacing the general parameter $\boldsymbol{\gamma}_j$ with
$\boldsymbol{\eta}_j$, and considering the difference
\begin{equation}
  \label{eq:delta}
  \boldsymbol{\delta}_j = \boldsymbol{\eta}_j^{(-j)} -
  \boldsymbol{\eta}_j^{(j)},
\end{equation}
where $\boldsymbol{\eta}_j^{(-j)}
\sim \pi(\boldsymbol{\eta}_j | \boldsymbol{y}_{-j})$ and $\boldsymbol{\eta}_j^{(j)}
\sim \pi(\boldsymbol{\eta}_j | \boldsymbol{y}_j)$. We then wish to test $H_0$:
$\boldsymbol{\delta}_j = \boldsymbol{0}$ for each group $j$. 
It is often reasonable to assume
multivariate normality of the posterior $\pi(\boldsymbol{\delta}_j |
\boldsymbol{y})$ (see the discussion in Section \ref{sec:disc} for
alternative approaches if we are not willing to assume approximate
posterior normality). If we denote the posterior expectation of
$\boldsymbol{\delta}_j$ by $\boldsymbol{\mu}(\boldsymbol{\delta}_j)$ and the posterior covariance by 
$\boldsymbol{\Sigma}(\boldsymbol{\delta}_j)$, the standardized discrepancy measure
\[
\Delta_j = 
\boldsymbol{\mu}(\boldsymbol{\delta}_j)^\top
\boldsymbol{\Sigma}(\boldsymbol{\delta}_j)^{-}
\boldsymbol{\mu}(\boldsymbol{\delta}_j)
\]
(where $\boldsymbol{\Sigma}(\boldsymbol{\delta}_j)^{-}$ is then a
Moore-Penrose generalized inverse of $\boldsymbol{\Sigma}(\boldsymbol{\delta}_j)$)
has a $\chi^2$-distribution
with $r=\text{Rank}(\boldsymbol{\Sigma}(\boldsymbol{\delta}_j))$
degrees of freedom under $H_0$. 
The discrepancy measure above is a generalization of the measure suggested by~\citet{gaasemyr2009} and
\citet[option (i) in Section 3.2.3 on page 384]{presanis2013}, extended to the case where $\boldsymbol{\Sigma}(\boldsymbol{\delta}_j)$ does not necessarily have full rank (typically, $\boldsymbol{\Sigma}(\boldsymbol{\delta}_j)$ will have less than full rank). 

For each group $j$, an estimate of  $\Delta_j$ is calculated by using two INLA runs: first, a ``between-group'' run is performed by removing (set to \texttt{NA}) the data in group $j$, providing an estimate for $\pi(\boldsymbol{\eta}_j | \boldsymbol{y}_{-j})$. Second, a ``within-group'' run is performed using only the data in group $j$, thus providing an estimated $\pi(\boldsymbol{\eta}_j | \boldsymbol{y}_{j})$. The ``shared parameter''/''cut'' problem described in item 3 above is dealt with by using an INLA feature allowing us to use the estimated posterior of the hyperparameters $\boldsymbol{\theta}$ from an INLA run as the \emph{prior} for a subsequent INLA run: we can then simply use the posterior of the hyperparameters from the ``between-group'' run as the prior for the ``within-group'' run. This corresponds exactly to~\citeauthor{plummer2015}'s \citeyearpar{plummer2015} recommended ``multiple imputation'' solution of the ``cut'' issue. 

Using the linear combination feature of INLA (see Section 4.4
of~\citet{martins2013}), estimates of the posterior means and
covariance matrices from both $\pi(\boldsymbol{\eta}_j |
\boldsymbol{y}_{-j})$ and $\pi(\boldsymbol{\eta}_j |
\boldsymbol{y}_{j})$ can be calculated quickly, and we can calculate the observed discrepancy 
\[
\widehat{\Delta}_j = 
\widehat{\boldsymbol{\mu}}(\boldsymbol{\delta}_j)^\top
\widehat{\boldsymbol{\Sigma}}(\boldsymbol{\delta}_j)^{-}
\widehat{\boldsymbol{\mu}}(\boldsymbol{\delta}_j)
\]
where
\begin{eqnarray*}
  \widehat{\boldsymbol{\mu}}(\boldsymbol{\delta}_j) &=& 
                                                        \widehat{\boldsymbol{\mu}}(\boldsymbol{\eta}_j^{(-j)}) -
                                                        \widehat{\boldsymbol{\mu}}(\boldsymbol{\eta}_j^{(j)}), \ \text{and}\\
  \widehat{\boldsymbol{\Sigma}}(\boldsymbol{\delta}_j) &=& 
                                                        \widehat{\boldsymbol{\Sigma}}(\boldsymbol{\eta}_j^{(-j)}) +
                                                        \widehat{\boldsymbol{\Sigma}}(\boldsymbol{\eta}_j^{(j)}).
\end{eqnarray*}

Finally, conflict $p$-values $p_1, \ldots,
p_J$ are then available as
\[
{p}_j = \text{Prob}(\chi_r^2 \geq \widehat{\Delta}_j),
\]
where $\chi_r^2$ denotes a chi-squared random variable with $r$
degrees of freedom.

\section{Implementation in R-INLA}
\label{sec:impl-r-inla}

The method has been implemented in the function \texttt{inla.cut()}
provided with the R-INLA package (see \url{http://r-inla.org}). The input to the function is an
\texttt{inla} object (the result of a previous call to the main
\texttt{inla()} function) and the name of the variable to group
by. The output is a vector containing conflict $p$-values for each
group. Thus, the usage is as follows:
\begin{verbatim}
inlares <- inla(...)
splitres <- inla.cut(inlares, groupvar)
\end{verbatim}
where \texttt{groupvar} is the grouping variable, and ``\texttt{...}''
denotes the arguments to the original \texttt{inla} call. See the
documentation of \texttt{inla.cut} (i.e., run \texttt{?inla.cut} in R
with R-INLA already installed)
for further details. Note that the \texttt{inla.cut} function is only
available in the ``testing'' version R-INLA; see
\url{http://r-inla.org/download} for details on how to download and
install this.

\section{Examples}
\label{sec:applications}

\subsection{Growth in rats}
\label{sec:rats}

Our first example is the ``growth in rats'' model considered in
Section 4.3 of \citet{presanis2013}, also analysed by
\citet{gelfand1990}. This data set consists of the weights $y_{ij}$
of rats $i=1,\ldots,30$ at time points $j=1,\ldots,5$ (corresponding
to ages 8, 15, 22, 29 and 36 days). This is modelled as
\begin{eqnarray*}
   y_{ij} &\sim& \text{N}(\mu_{ij}, \tau),\\
  \mu_{ij} &=& \beta_0 + \beta_1 t_j + \psi_{i0} + \psi_{i1} t_1,\\
   \boldsymbol{\psi}_i &\sim& \text{MVN}_2(\mathbf{0}, \boldsymbol{\Omega}),
\end{eqnarray*}
where ``MVN$_2$'' denotes the bivariate normal distribution, $\boldsymbol{\psi}_i = (\psi_{i0}, \psi_{i1})'$, 
the fixed effects $\beta_0$ and $\beta_1$ are given independent
$N(0, 10^{-6})$ priors,  the data precision $\tau$ is
given a $\Gamma(10^{-3}, 10^{-3})$ prior, and the precision matrix $\boldsymbol{\Omega}$ of
the random effects is given the prior
\[
\boldsymbol{\Omega} \sim \text{Wishart}\left(
  \begin{pmatrix}
    200 & 0\\
    0 & 0.2
  \end{pmatrix}
, 2\right).
\]

\begin{table}[!ht]
\centering
\begin{tabular}{rrr}
  \hline
 & MCMC & INLA \\ 
  \hline
1 & 0.97 & 0.96 \\ 
  2 & 0.06 & 0.06 \\ 
  3 & 0.74 & 0.74 \\ 
  4 & 0.11 & 0.11 \\ 
  5 & 0.18 & 0.17 \\ 
  6 & 0.83 & 0.81 \\ 
  7 & 0.62 & 0.59 \\ 
  8 & 0.86 & 0.86 \\ 
  9 & 0.0025 & 0.0026 \\ 
  10 & 0.21 & 0.21 \\ 
  11 & 0.32 & 0.32 \\ 
  12 & 0.51 & 0.49 \\ 
  13 & 1.00 & 1.00 \\ 
  14 & 0.16 & 0.15 \\ 
  15 & 0.07 & 0.08 \\ 
  16 & 0.68 & 0.68 \\ 
  17 & 0.58 & 0.56 \\ 
  18 & 0.69 & 0.70 \\ 
  19 & 0.72 & 0.73 \\ 
  20 & 0.95 & 0.95 \\ 
  21 & 0.87 & 0.87 \\ 
  22 & 0.45 & 0.45 \\ 
  23 & 0.53 & 0.50 \\ 
  24 & 0.63 & 0.63 \\ 
  25 & 0.03 & 0.02 \\ 
  26 & 0.65 & 0.64 \\ 
  27 & 0.26 & 0.26 \\ 
  28 & 0.64 & 0.63 \\ 
  29 & 0.18 & 0.16 \\ 
  30 & 0.99 & 0.99 \\ 
   \hline
\end{tabular}
\caption{$P$-values from the rats example.} 
\label{tab:ratp}
\end{table}

For this model, it seems natural to ask whether any of the rats are
``outliers'' in the sense of the model being unsuitable for the
particular rat. Using the node-splitting approach, we can address this
question, giving a $p$-value corresponding to a test of model adequacy for
each rat.
The resulting $p$-values are shown in Table~\ref{tab:ratp}, where we
also compare to the result from a long MCMC run using
JAGS~\citep{plummer2016} (details not shown).

\begin{figure}
\includegraphics[width=\maxwidth]{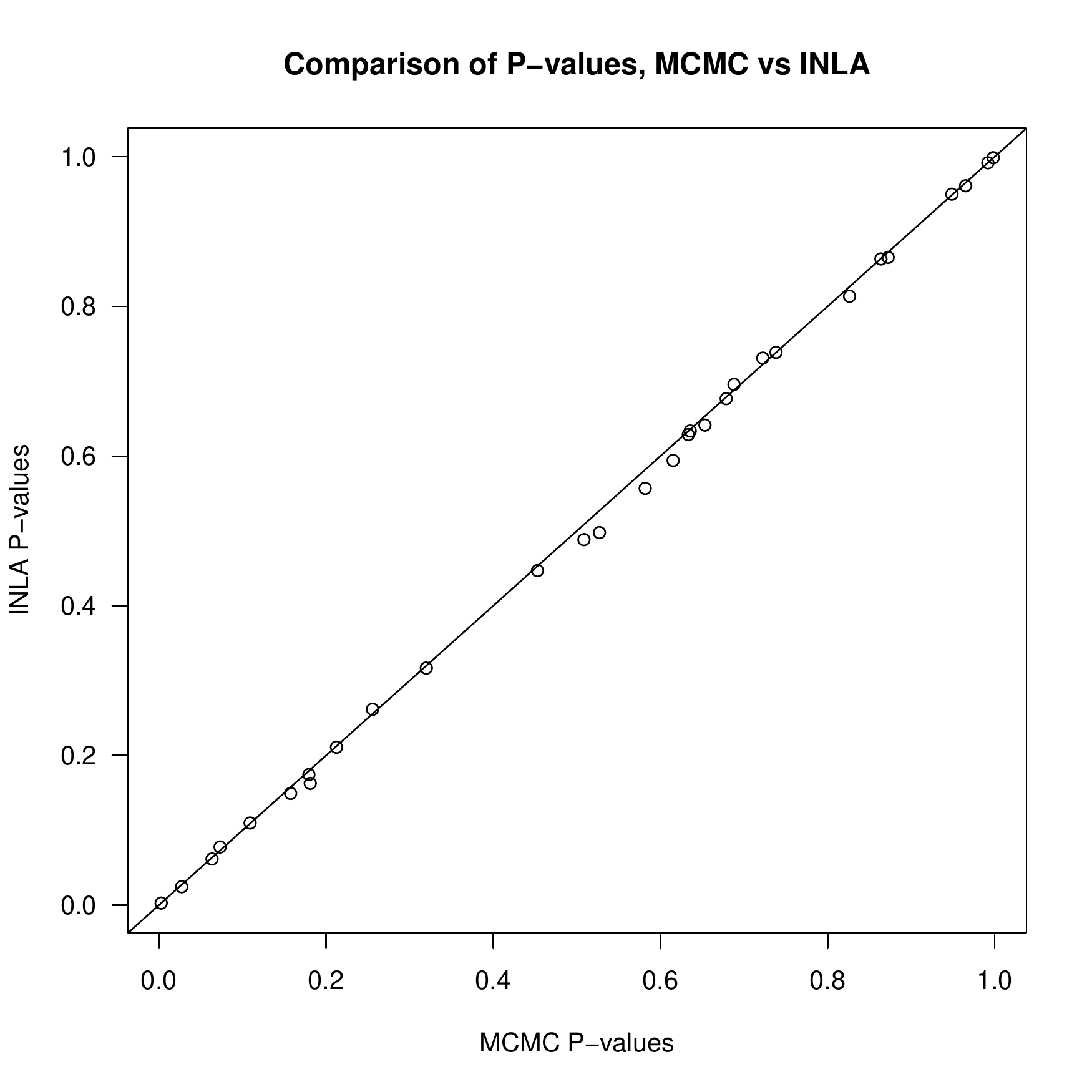} 
\caption{Scatterplot of MCMC vs.~INLA $p$-values.\label{fig:scatter}}
\end{figure}

\begin{figure}[htbp]
  \centering
  \includegraphics[width=\linewidth]{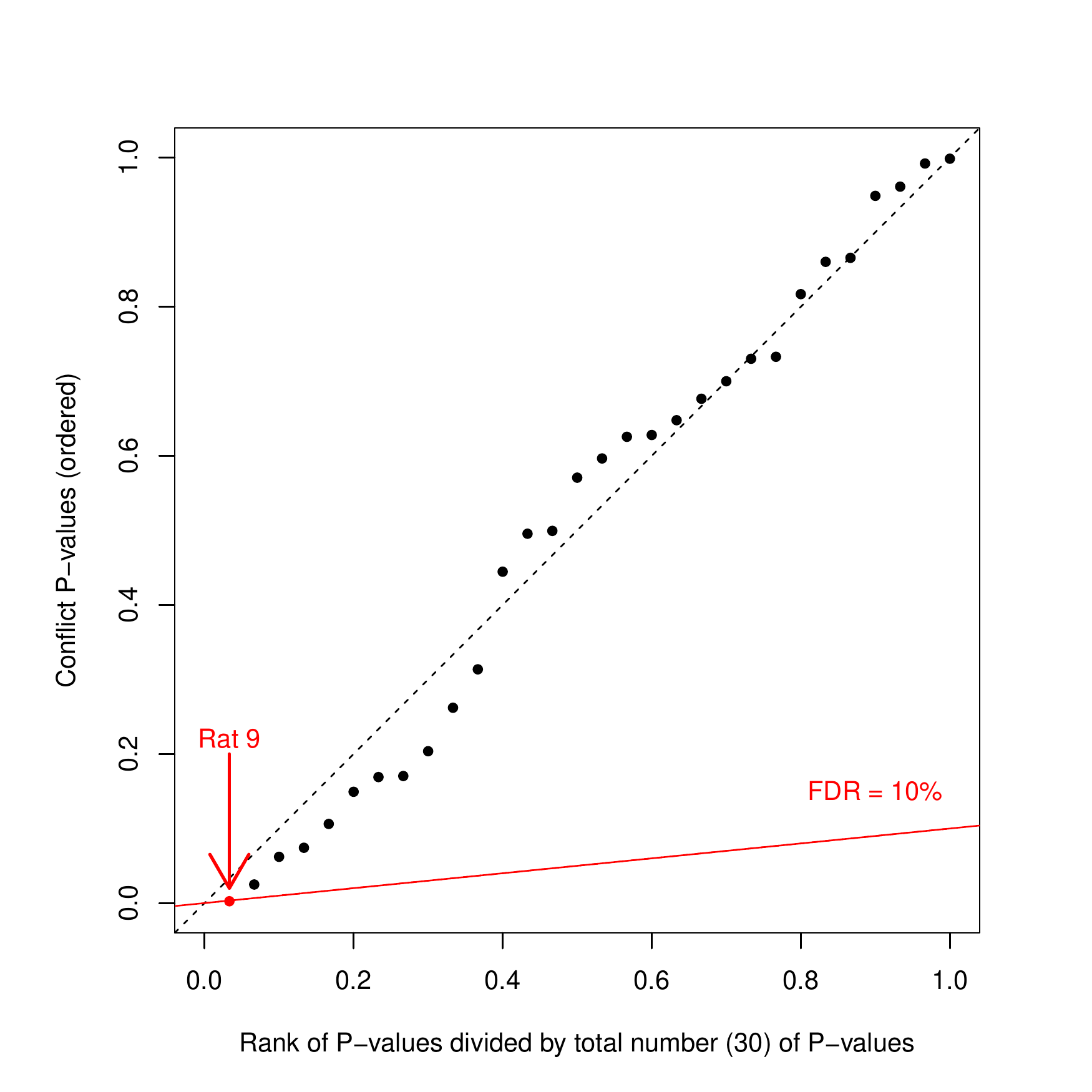}
  \caption{$P$-values from rat growth data. The Benjamini and Hochberg
    criterion is superimposed as a red line, identifying
    rat 9 as being divergent. \label{fig:ratsp}}
\end{figure}

Figure \ref{fig:scatter} shows a scatterplot of MCMC versus INLA
$p$-values. There is not any substantial difference between the MCMC
and INLA results.  Figure \ref{fig:ratsp} shows the empirical
distribution of the resulting $p$-values. Rat number 9 is identified
as an ``outlier'' ($p$=0.0021) at a false discovery
rate~\citep{benjamini1995} of 10\%.

The total INLA run time for this example was 8.3 seconds, with the
initial INLA run taking 2.5 seconds and the node-splitting using
\texttt{inla.cut} (see Section~\ref{sec:impl-r-inla}) taking 5.8
seconds. 
A machine with 12 Intel Xeon X5680 3.33 GHz CPUs and 99 GB RAM was used.
The MCMC run time using
the same hardware, running JAGS and two chains of 200,000 iterations
each (discarding the
first 100,000 iterations), was 34 minutes, so INLA was nearly 250 times
faster than MCMC for this example. While it is of course difficult to
know for how long it is necessary to run an MCMC algorithm, investigation of convergence diagnostics
from the MCMC run indicates that the stated iterations/burn-in is not
excessive.

\subsection{Low birth weight in Georgia}
In this example, we study the ``Georgia low birth weight'' disease
mapping data set
from Section 11.2.1 of~\citet{lawson2013}. Different analyses of these data using
INLA are described in~\citet{blangiardo2013}. The data consist of
counts of children born with low birth weight (defined as less than
2500 g) for the 159 counties of the state of Georgia, USA, for the 11
years 2000-2010. Let $y_{ij}$ be the count for county
$i=1,\ldots,159$, year $j=1,\ldots,11$, and let $t_j = j - 5$ be the centered
year index. A possible
model is then
\begin{eqnarray}
   y_{ij} &\sim& \text{Poisson}(E_{ij}\exp(\eta_{ij})), \nonumber \\
  \eta_{ij} &=& \mu + u_i + v_i + \beta t_j, \label{eq:spatial.model}
\end{eqnarray}
where $E_{ij}$ is the total number of births, and $u_i$ and $v_i$
are together given a Besag-York-Mollie (BYM)
model~\citep{besag1991}. In the BYM model, the spatially structured random effect $u_i$
is given a intrinsic conditionally independent autoregressive
structure (iCAR), with $u_i | u_{j \neq i} \sim N(n_i^{-1}
\sum_{\mathcal N_i} u_j, n_i^{-1} \sigma_v^2)$, where $n_i$ is the number of
neighbors (sharing a common border) with county $i$, and $\mathcal
N_i$ denotes the set of neighbors of $i$.  The $v_i$ are residual
unstructured effects, independent and identically distributed $N(0, \sigma_u)$. The INLA default low-informative
priors are defined on the hyperparameters $\theta_u = \log \sigma_u^{-2}$
and $\theta_v = \log \sigma_v^{-2}$: both $\theta_u$ and
$\theta_v$ are given Gamma$(1,0.0005)$ priors.

For this model, traditional model-check diagnostics are either on the
(county, year) level (checking the fit for each individual count
$y_{ij}$) or on the overall model (as we could do with standard
goodness-of-fit tests). However, using the node-splitting methodology
we can also either
\begin{enumerate}
\item check the model adequacy by county, or
\item check the model adequacy by year,
\end{enumerate}
simply by grouping by county or year, respectively.

\begin{figure}[htbp]
  \centering
  \includegraphics[width=\linewidth]{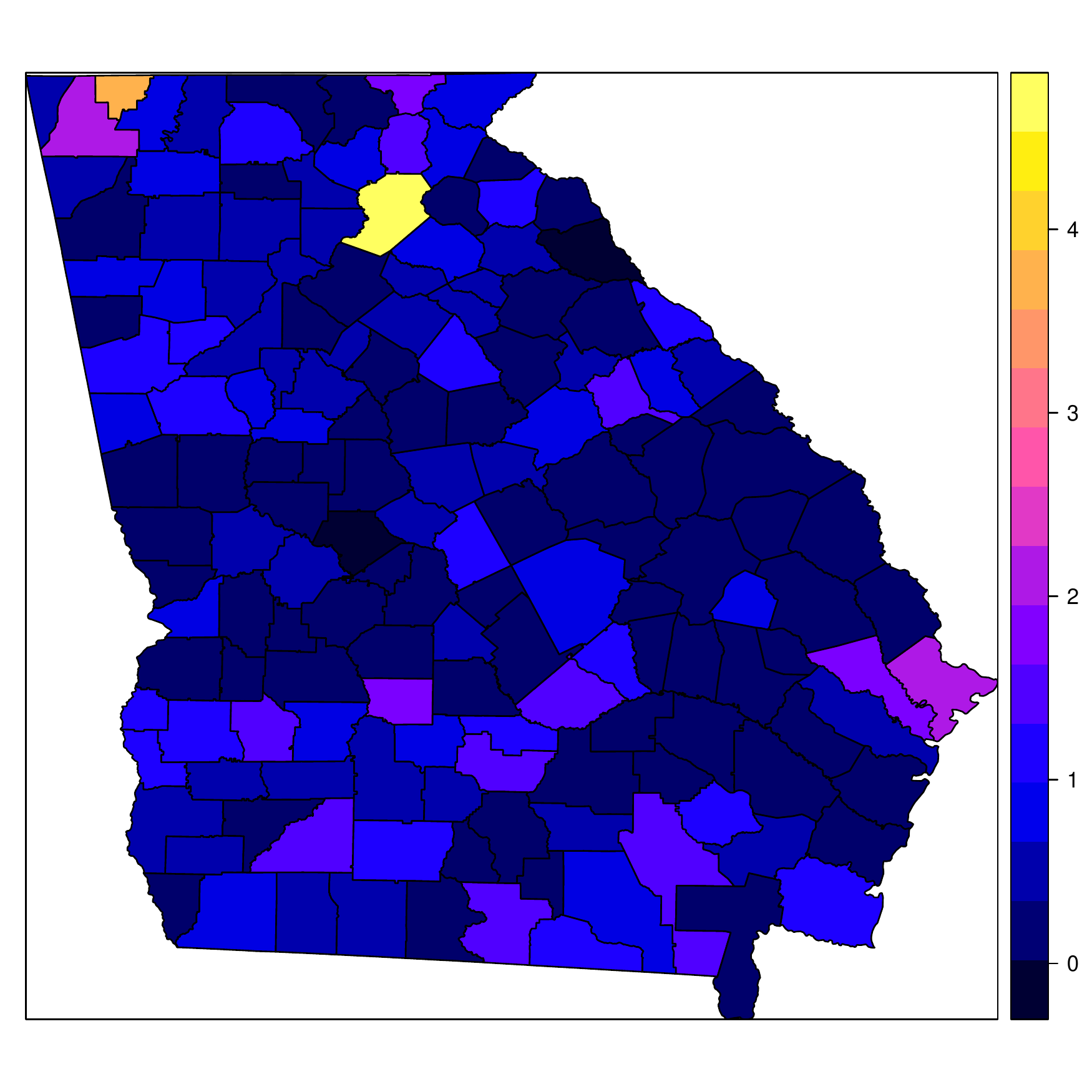}
  \caption{Map of $-\log_{10}$ of $p$-values from spatial split of low
    birth weight model.\label{fig:lbw_allp}}
\end{figure}

\begin{figure}[htbp]
  \centering
  \includegraphics[width=\linewidth]{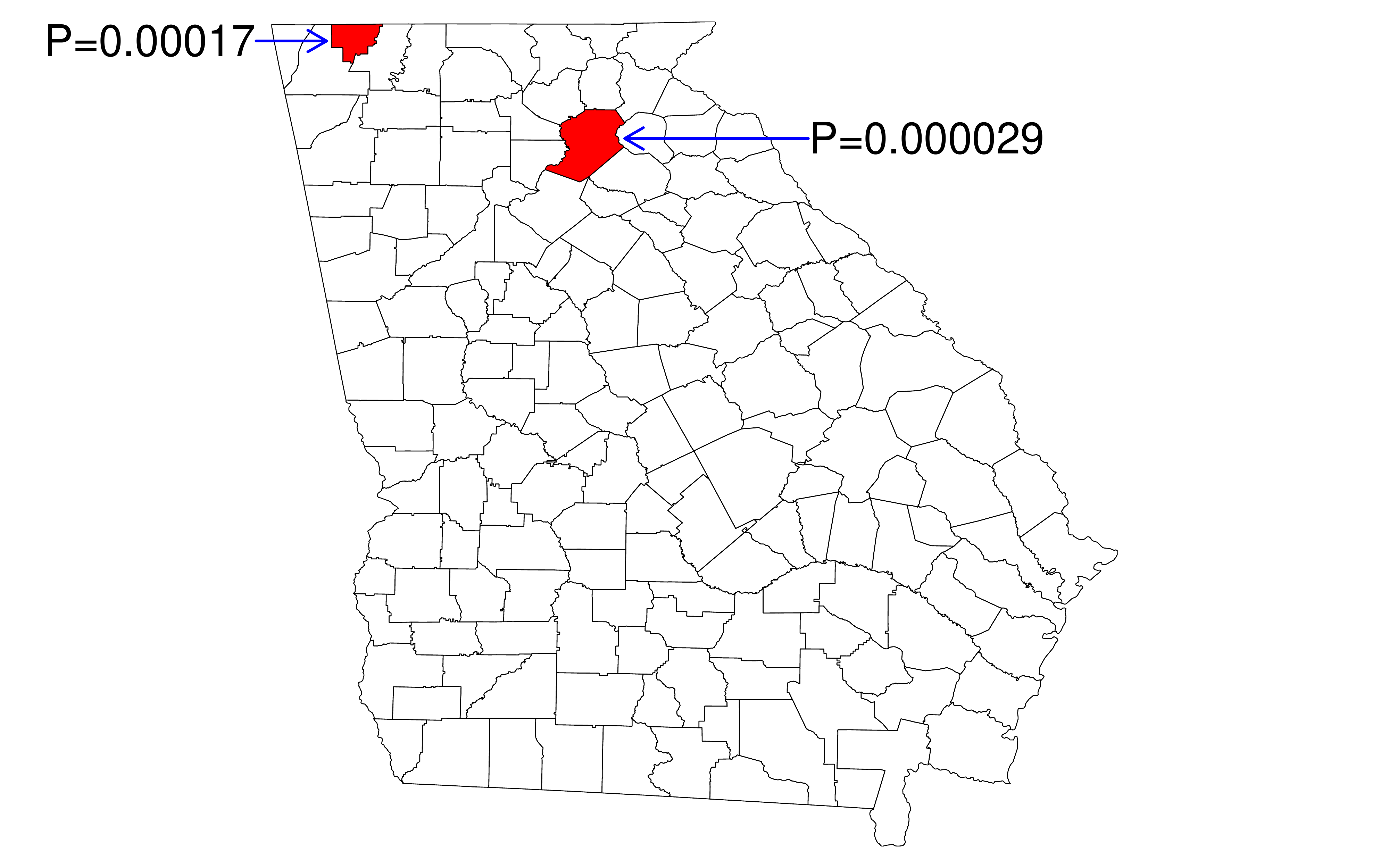}
  \caption{The two ``outlier'' counties identified as not conforming to the model.\label{fig:lbw_outliers}}
\end{figure}

Figures \ref{fig:lbw_allp} and \ref{fig:lbw_outliers} show the results
from the (spatial) split by county. Two counties are identified as divergent at a 10\% false discovery rate (FDR) level: Hall and Catoosa county, with 
$p$-values of 0.000029 and 0.00017, respectively.

\begin{figure}[htbp]
  \centering
  \includegraphics[width=\linewidth]{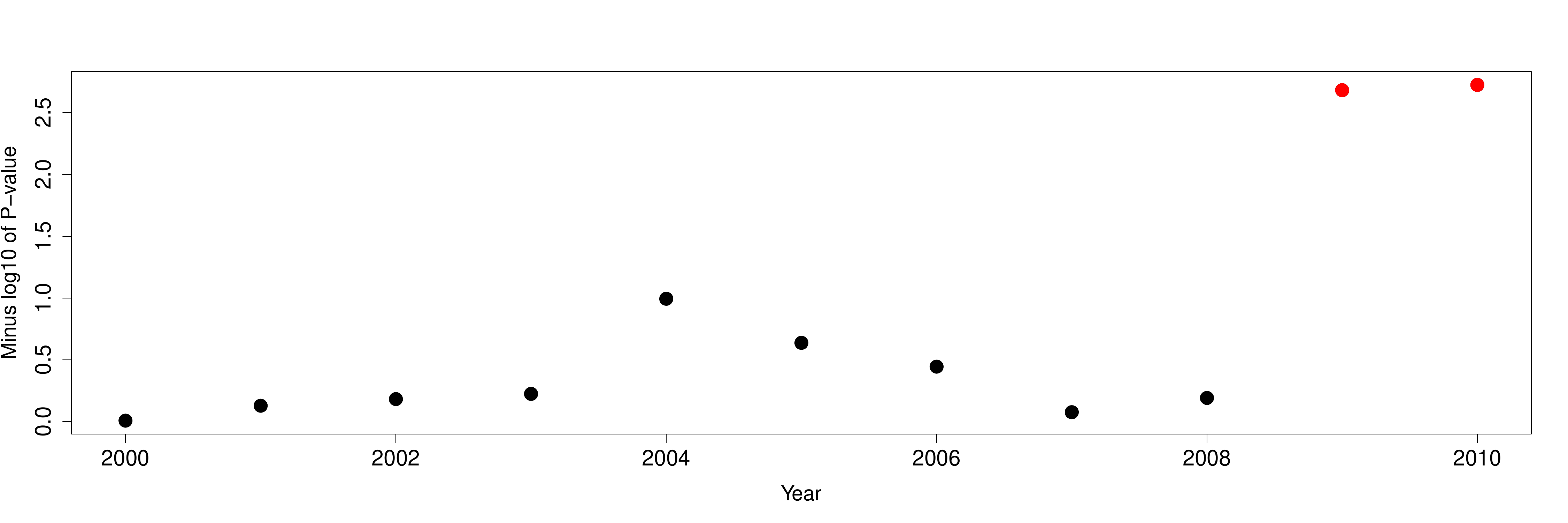}
  \caption{Time series of $-\log_{10}$ of $p$-values from temporal (yearly) split of low
    birth weight model. The two last years
    (marked in red) are identified as outliers.\label{fig:lbw_yearsplit_p}}
\end{figure}

It is also interesting to study the model adequacy by year: Are any of
the years (from 2000-2010) identified as outliers?
Figure~\ref{fig:lbw_yearsplit_p} shows the results ($-\log_{10}$ of
$p$-values) from node-split when we group by year. We see that the last two years, 2009 and
2010, are identified as divergent, with $p$-values of 0.0022 and 0.0020,
respectively. This suggests that the assumption of a linear time trend in 
model~\eqref{eq:spatial.model} needs some amendment. 

The run time (using the same hardware as in Section~\ref{sec:rats})
was approximately four seconds for the initial INLA run, 30 seconds for
the split by year, and approximately five minutes
for the split by county. We did not implement any MCMC algorithm for
this example.

\section{Discussion}
\label{sec:disc}
Although our aim has been to provide a general, easy-to-use
implementation, there are some limitations. First, we currently assume that the grouping is ``linear'' in
the sense of always comparing group $i$, say, with all remaining
groups (``$-i$''). To implement the method for the case of network
meta-analysis, it would be necessary to extend the implementation to the
more general case where group $i$ might be compared with some, but not
all, of the remaining groups, and the list of ``remaining groups''
might change arbitrarily with $i$. Second, we only consider conflict
detection based on the linear predictor of the model; in some cases it
might be necessary to consider other parameters. However, this would
make it more challenging to provide a general implementation. Third,
we restrict to the class of latent Gaussian models, but this is of
course more a limitation of the INLA methodology itself and not of
the model criticism method as such. Finally, as described above, the
method (as currently implemented) relies on an assumption of
approximate multivariate normality of the
posterior linear predictor. As discussed in Section 3.2.3 on page 384
of~\citet{presanis2013}, there are ways to avoid this assumption, at the
cost of a higher computational load. For example, if we only want to
assume a symmetric and unimodal (but not necessarily normal) posterior
density, an alternative test can be devised based on sampling from the
posterior and using Mahalanobis
distances. Since R-INLA provides functionality for quickly producing
independent samples
from the approximated joint posterior, the Mahalanobis distance based
approach could potentially be implemented within our
approach. However, in our experience the posterior distribution of
the linear predictor is nearly always sufficiently close to normal; thus,
our judgment has been that using the Mahalanobis-based p-values was not
worth the extra computational effort, and it is not currently
implemented.
 Nevertheless, if there is a popular
demand, this functionality
might be added in a future version of the software.

The method is cross-validatory in nature and avoids ``using the data
twice'', a well-known problem with posterior predictive $p$-values and
related methods~\cite{bayarri2007}. Nevertheless, it would be a useful
topic for further study to consider both the power of our conflict
detection tests, as well as the ``calibration'' in the sense that
$p$-values from a true model should be uniformly
distributed (see~\citet{gaasemyr2016} for some recent work on the
calibration of node level conflict measures). 
In particular, in some cases the method may not detect a "practically significant" conflict due
to insufficient power. Power (e.g.~with FDR thresholding) has not yet
been investigated systematically so far, and this could potentially be carried out 
in a simulation study with INLA, whereas MCMC would be too slow to do
this. Simulation studies could also be useful for assessing the
calibration issue mentioned above, i.e.~whether $p$-values are really
always uniform under no conflict.

To account for the multiple testing issues generated by
considering many groups and performing a test for each group, we have
simply used FDR. FDR correction provides a
convenient and well-understood methodology for dealing with multiple
testing. However, it could still potentially be useful to employ
multiple testing methodology that is more specifically geared to 
the group-wise model checking tests used here. See~\cite{presanis2017}
for some work in this direction.

Finally, it could be useful to provide graphical or more general
numerical measures of fit in addition to the conflict $p$-values, in
order to obtain further insight into the reasons for any discovered lack
of fit, for example, using the approach of~\citet{scheel2011}. We leave this as an interesting topic for further work.

\section*{Acknowledgments}

We thank Anne Presanis for providing code and
documentation for an MCMC implementation of the ``rats'' example. Thanks are also due to Lorenz Wernisch and Robert Goudie for
helpful comments.

\bibliographystyle{apalike}

\end{document}